\documentclass{article}

\usepackage{subcaption}
\usepackage{hyperref}
\usepackage{graphicx}
\hypersetup{breaklinks=true}

\title{Towards Citizen Science for Smart Cities: 
A Framework for a Collaborative Game of Bird Call Recognition Based on Internet of Sound Practices
}

\author{
  Emmanuel Rovithis\\
  \texttt{emrovithis@ionio.gr}
  \and
   Nikolaos Moustakas\\
  \texttt{a11mous@ionio.gr}
   \and
   Konstantinos Vogklis\\
  \texttt{voglis@ionio.gr}
   \and
   Konstantinos  Drossos\\
  \texttt{constantinos.drossos@tuni.fi}
   \and
   Andreas Floros\\
  \texttt{floros@ionio.gr}
}

\begin{document}

\maketitle

\begin{abstract}
Citizen Science aims to engage people in research activities on important issues related to their well-being. Smart Cities aim to provide them with services that improve the quality of their life. Both concepts have seen significant growth in the last years, and can be further enhanced by combining their purposes with IoT technologies that allow for dynamic and large-scale communication and interaction. However, exciting and retaining the interest of participants is a key factor for such initiatives. In this paper we suggest that engagement in Citizen Science projects applied on Smart Cities infrastructure can be enhanced through contextual and structural game elements realized through augmented audio interactive mechanisms. Our inter-disciplinary framework is described through the paradigm of a collaborative bird call recognition game, in which users collect and submit audio data, which are then classified and used for augmenting physical space with virtual soundscape maps. We discuss the Playful Learning, Internet of Audio Things, and Bird Monitoring principles that shaped the design of our paradigm, and analyze its potential technical implementation.        
\end{abstract}

\section{INTRODUCTION}
\label{sec:introduction}

The concept of Smart Cities describes urban environments enriched with interaction modalities towards the improvement of city functioning and the life of its inhabitants \cite{batty2012smart}. Aiming to align the technological attainments of the digital era with the urban fabric of the physical world, Smart Cities utilize Information and Communication Technologies (ICT), such as mobile devices, embedded sensors, and data collection and analysis tools, and seamlessly integrate them in traditional infrastructure. Thus, the physical environment is transformed into a dynamic source of information, an "intelligent living space", which, based on the adoption of networking advances, provides citizens with the tools, resources, and services to exploit the benefits of this data flow \cite{urbieta2017adaptive}. The realization of intelligent systems to be embedded into Smart City environments can be broken down into three axes: i) the Internet of Things (IoT) facilitating the inter-connectivity of physical and virtual devices through communication protocols, ii) the Internet of Services (IoS) comprised of the amalgamation of different applications into explicable services, and iii) the Internet of People (IoP) encompassing the interactions between the citizens, who are ultimately the intended users of the system \cite{hernandez2011smart}. Online, mobile, and augmented technologies provide the ground, on which user-centred services for data tracking, monitoring, and processing will flourish, yet, the defining factor for Smart Cities to become truly beneficial intelligent ecosystems is to correlate the appropriate tools, techniques, methods, and organizational structures that will engage the community towards the improvement of the quality of life \cite{batty2012smart}. In exciting the citizens' creativity and promoting their collaboration Smart Cities become innovation ecosystems capable of finding solutions to real-world problems \cite{schaffers2011smart}. To that scope, Smart Cities require applications that facilitate large-scale participatory projects, in which emphasis will be placed on the coordination of end-users towards dealing with the targeted issues \cite{schaffers2011smart}. 

The concept of Citizen Science describes projects, in which people volunteer to contribute to a scientific enquiry by gathering and managing information. The advent of the 21st century has signaled an unprecedented growth of Citizen Science initiatives bringing scientists and the public together with the aim of raising awareness and finding solutions about social and environmental issues including public health, social trends, human mobility, technological impact, information exchange, natural resources, environmental conditions, and species protection. Volunteers can now quickly locate a citizen science project on a subject of their interest and easily join its active community, whereas advances in human-computer interaction have extended the access to such projects to groups that could not previously be reached \cite{bonney2014next}. Thus, citizens can take on a vital role in research processes that may vary from simply providing experts with the necessary information to consulting the responsible authorities or even participating in making and implementing decisions \cite{conrad2011review}. Such endeavours require from an organizational perspective the coordination of citizens, government institutions, agents of industry and academia, and local community groups. In terms of content, they must adhere to basic scientific principles, such as well designed data collection and validation methods, explicit instructions and research questions, and feedback as a reward for participation \cite{silvertown2009new}. 

We believe that Smart Cities can host large-scale participatory Citizen Science projects in a mutually beneficial relationship, in which Smart Cities provide the technological infrastructure and Citizen Science the activities targeting the citizens' well-being. However, the design of such an endeavour must address important challenges related to participants' engagement and cognition. Regarding the former, volunteer dropout has been identified as one of the salient factors that impact the organizational consistency of Citizen Science \cite{conrad2011review, hand2010citizen}. Regarding the latter, undertaking and accomplishing tasks in a Citizen Science context does not necessarily extend beyond the acquisition of knowledge to a deeper understanding of the scientific process \cite{bonney2016can}. So far, suggestions for exciting and retaining the interest of the citizen scientists include providing positive reinforcement, and matching the tasks to the personal skills of the volunteers \cite{conrad2011review}. Cognitive impact has been approached mostly through student assessments in curriculum-based projects presenting limited evidence that participation enhances scientific knowledge and public awareness, and thus more evaluations are deemed necessary for extracting solid conclusions \cite{bonney2016can}. In this paper we suggest that motivation and understanding can be enhanced through an inter-disciplinary approach that combines structural and contextual game elements with Internet of Audio Thing (IoAuT) technologies \cite{turchet2020internet} to realize Citizen Science projects in Smart Cities environments. We describe our proposed framework for developing a Citizen Science for Smart Cities project through the paradigm of a bird call recognition augmented reality audio-based game. 

The rest of the paper is organized as follows. In Section~\ref{sec:theoretical} are discussed the principles underlying the paradigm’s design in terms of its playful learning, audio interaction, and bird recognition aspects, Section~\ref{sec:framework} presents the conceptual and technical structure of the paradigm and Section~\ref{sec:technical_implementation} describes its technical implementation. Finally, Section~\ref{sec:conclusions} concludes the paper.

\section{DESIGN PRINCIPLES}
\label{sec:theoretical}
\subsection{Playful Learning}

Playful Learning descries the incorporation of game elements into non-game learning environments \cite{plass2015foundations}.The ability to motivate players is the most frequently cited characteristic of games related to knowledge construction \cite{costa2016review}. By utilizing a variety of interaction mechanisms games create the conditions for competition, cooperation, exploration, and reflection, and engage participants in immersive experiences \cite{dicheva2015gamification}. Aiming to investigate the connection of motivation and engagement to the learning outcomes researchers have intensified their efforts in the last decade, whereas educators have been drawing upon the results to systematically use game-based learning practices in their classroom \cite{costa2016review}. Non-schooling environments have been also following this trend: museums, libraries, corporations, and government agencies have been integrating game elements in personal or collective activities as the means to enrich users’ experience and enhance their construction of knowledge. However, simply adding a leaderboard system based on points of progress, may have negative effects, since players with low scores could become frustrated and lose their interest in the competition \cite{reiners2014chaos}. The same applies to game systems that address the broader community: stereotypical approaches will not necessarily result in increasing and sustaining participation \cite{thiel2016playing}. Therefore, careful planning of game elements integrated into non-game systems is needed to ensure motivation at all times of the process. 

Large-scale participatory Citizen Science projects require attention, coordination, cooperation, and commitment. The few cases, in which Citizen Science was organized in the form of a game, have delivered positive results: providing users with a playful interface and allowing them to collaborate with or compete against each other towards a common goal resulted in users coming up with novel ideas \cite{hand2010citizen}. Another approach refers to Citizen Science projects, which are embedded in the form of mini-games within larger sand-box game environments, i.e. environments, in which players have freedom of action that is not restricted by a strict linear narrative. In the case of \cite{CCP} players completing various stages of the mini-games are rewarded with in-game prizes.

Besides the motivational function, there are other game elements that can be useful to Citizen Science. In order for Citizen Science to produce an output of equal-to-expert quality, the participants need guidance through protocols, training, and oversight \cite{bonney2014next}. A game’s rules, tutorial, and feedback can address these issues respectively, whereas the addition of a compelling narrative can enhance immersion in the experience. A final issue that we considered is the link of high motivation and engagement to the intended learning outcomes. Drawing upon modern learning theories from the field of Education, including Problem-based Learning \cite{boud1997challenge}, i.e. learning from the process of striving toward the resolution of a problem, Constructivist Learning \cite{duffy1991constructivism}, i.e. learning from the process of interacting with the environment, and Experiential Learning \cite{kolb2014experiential}, i.e. learning from the process of reflecting on one’s experience, provides the theoretical basis for realizing meaningful learning environments \cite{kiili2005educational}. Yet, researchers stress the need for stronger evidence, before game mechanisms aiming at motivation and immersion are systematically used as the means to achieve the learning objectives \cite{perrotta2013game, hamari2016challenging}, a need that large-scale participatory projects can address. 

\subsection{Internet of Audio Things}

IoAuT is an emerging field that refers to embedding computing devices in physical objects towards the reception, processing, and transmission of audio information \cite{turchet2020internet}. It comprises different types of audio collectors, processors and transmitters, and facilitates their integration, local and remote accessibility, and multi-directional communication. Despite the plethora of Smart City initiatives and the need for utilizing state of the art Human-Computer Interaction (HCI) technologies to realize new forms of participation, most approaches have focused on data visualisation techniques and mostly neglected the acoustic aspect of the urban environment \cite{batty2012smart,sarmento2020musical}. Existing Smart City applications of distributing information through the auditory channel include the generation of sound content based on urban related data in order to increase users’ awareness about their city environment \cite{sarmento2020musical,drossos:2012:am}, the generation of visual maps based on the perceptual attributes of submitted recordings for monitoring and managing the urban acoustic environment \cite{kang2018model}, and the augmentation of public spaces with audio besides visual information for engaging the audience in social experiences \cite{nikolic2020designing}. Furthermore, Wireless Acoustic Sensor Networks can be used for the surveillance and analysis of acoustic scenes, urban noise pollution, environmental anomalies, and wildlife \cite{turchet2020internet}. 

In our paradigm design we focused on three specific aspects of IoAuT: i) collecting and submitting audio data for analysis, ii) generating a soundscape map, and iii) augmenting physical space with virtual audio components for navigation and interaction within the environment. Focusing on the latter, Augmented Reality Audio (ARA) systems have been applied for well-being purposes by acoustically enriching the working environment of employees \cite{mynatt1997audio}, indicating the location of security threats \cite{sundareswaran20033d}, realizing non-visual spatial mappings for navigation \cite{holland2002audiogps}, aurally signalling touristic points of interest \cite{mcgookin2009audio}, assigning audio recordings to locations of cultural importance \cite{reid2005parallel}, and aurally signifying city facilities for urban exploration \cite{blum2011s}. Interaction in these implementations can be characterised as passive, i.e. users of the system essentially trigger sound events through their position and movement in the augmented space. However, more active modes of interaction can enhance users’ communication in competitive or collaborative contexts. In \cite{rovithis2019audio} a positive connection was shown between challenging mechanics requiring the performance of gestures with the satisfaction from the experience. In \cite{moustakas2011interactive} the behavior of the virtual sound sources that players need to locate is controlled by the movement of other antagonizing players, whereas in \cite{pellerin2009soundpark} players take up different roles and need to coordinate their actions in the augmented space to achieve the game goal. Sound recognition and audio based analytics~\cite{drossos:ijcnn:2020,drossos:icassp_b:2020} can further expand the possibilities for interaction by advancing the responsiveness between the natural and the virtual acoustic environment~\cite{pulkki2018parametric}, whereas user experience improvement techniques from the wider frame of AR can be utilized to enhance the system’s context-awareness \cite{aliprantis2019survey}. 

\subsection{Bird Monitoring}

The third field that we drew upon for designing our paradigm relates to Bird Monitoring. Bird related ecological projects usually fall into three categories: i)inventory, ii) monitoring, and iii) research. Inventory projects aim to generate a list of species by identifying birds by visual observation and/or call. Monitoring projects involve recording birds in a region or study site for a period of time. Such projects use geolocation information to pinpoint found birds on Geographic Information System (GIS) overlays. Research projects require experts to formalize and investigate a hypothesis about bird behaviour.

One of the leading active projects in collaborative Bird Monitoring is eBird, a project of the Cornell Lab of Ornithology~\cite{sullivan2009ebird,sullivan2014ebird}. eBird  evolved from a basic Citizen Science project into a collective enterprise through the novel approach of developing cooperative partnerships among experts in a wide range of fields including computer scientists, biologists, and data administrators. eBird data are overlaid on global GIS maps. They are openly available and constitute a major source of biodiversity data, increasing expert knowledge on the dynamics of bird species distributions and aiding the conservation of birds and their habitats. The project is now targeted only for expert biologists and bird watchers and relies on their expertise to deliver up-to-date results about bird populations. It lacks two major features that we are planning to address through our proposed methodological framework: i) the involvement of non-expert users, and ii) game-based modes of interaction.

\section{PROPOSED FRAMEWORK}
\label{sec:framework}

\subsection{Scenario Design}

Our paradigm consists of three stages. In the first stage, users undertake the task of collecting and submitting audio material for classification. They select among available quests, which focus on specific birds, and then walk around the city to spot and record them. Recordings are performed via their mobile device. Once a recording is submitted to the system, it is recognized and classified into the respective bird species. Users are provided with information about the bird they recorded. Their total progress in gathering data rewards them with progress points and badges, which grant access to a data bank with sounds of each bird for further study, and unlock more complex quests with combinations of birds to be collected within limited time-frames. 

In the second stage, meta-data related to the recording including date and location are used for the creation of a soundscape representing bird activity in the urban environment. Data sonification and sound spatialization techniques are used to express the reported aspects of the aerial fauna: panning hints at the location of birds, spectral frequency at their migratory mobility, playback volume at their proximity to the user, and playback rate at the amount of the submitted recordings. The soundscape is dynamically shaped according to users' position in space. Users can also control the parameter of time, allowing them to monitor the birds' behaviour through long time periods.

In the third stage, the virtual soundscape is superimposed on physical space, allowing users to immerse themselves in the augmented environment. By turning their mobile device to different directions they can scan their surroundings and locate points of interest defined through a customizable set of filters. Once navigated to a point of interest, users can use the tilt operation of the device to receive specialized audio information regarding the characteristics of the targeted birds. In essence, the first stage aims at creating a soundscape map to express bird activity, the second stage at allowing virtual navigation through the soundscape in the axis of time, and the third stage at allowing augmented navigation in the axis of space. 

\subsection{Architecture Design}

\begin{figure}[!h]
\centering
\includegraphics[width=10cm]{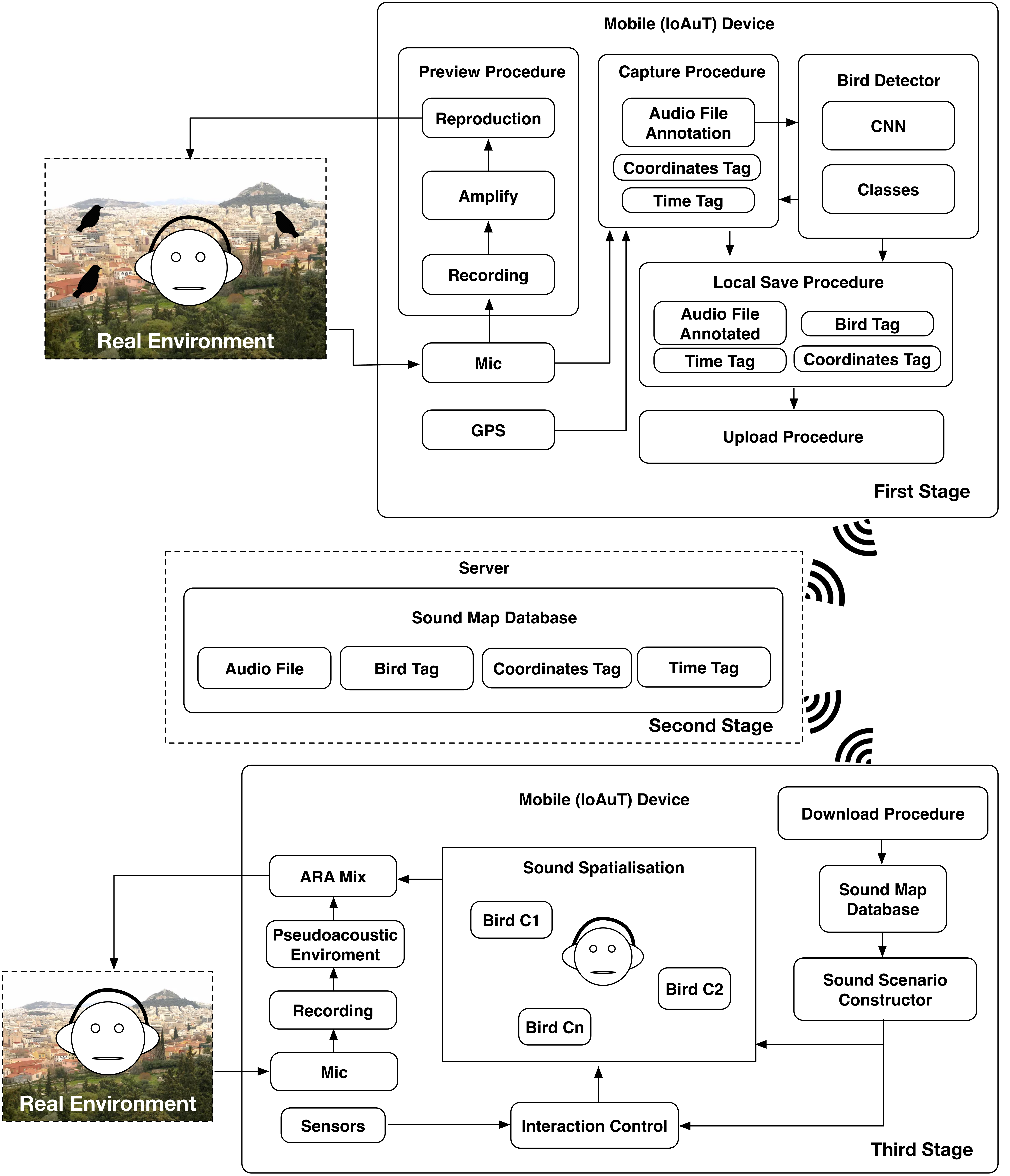}
\caption{The Concept Architecture}
\label{fig:TheConceptArchitecture}
\end{figure}

The design of our paradigm's architecture is based on the three stages defined in the scenario, and involves a three-layer IoAuT setup. The Sensing Layer includes the sensors, the recording and the playback module in the user's mobile device, which allow for producing audio content and analyzing phenomena associated with auditory events \ref{fig:TheConceptArchitecture}. The Network Layer is responsible for data transfer from the Sensing Layer, and the Application Layer includes the web services and the virtual soundscape construction module.

Focusing on exploring the ARA environment, the architecture design segments the concept into two primary modes. The first mode is designed to facilitate passive interaction, while users walk through the real environment. After the desired filters are set through the menu of the mobile app, the sound monitoring mechanism reproduces in real-time the real acoustic environment, and the playback mechanism plays back the captured sound, when its source is in the user's proximity. All audio components are mixed together and delivered through the audio headset. An amplification coefficient, which is adjustable by the user, is applied to audio capture, for improving the birds sound intelligibility. Once users hear something of interest, they can enter the second mode and actively search the dynamically generated sound map of stored birds’ sounds to locate points of interest and interact with them.  
 
Once the preview mechanism is derived using the above procedure, the audio recording can be implemented into the existing capture procedure model. The user then responds to this procedure by annotating the part of the waveform, which contains the bird sound. A Convolutional Neural Network (CNN) model, either embedded in the mobile device or situated on a back-end server, checks to recognize specific bird classes. If the model classifies the annotating sound to a specific class, then the local save procedure is enabled. More specifically, the following is stored: i) the annotation of the audio file, ii) the tag of the bird class, iii) the tag of the time of the event, including date, year, and hour, and iv) the GPS coordinates that are captured using the GPS features of the mobile device. As soon as the local saved data are ready, the final upload procedure to the web server can be made.  

The Application Layer realizes the third stage, which is organized in a client-server architecture. The mobile app, as the client, asks for a sound scene by sending the phone position, in order to download the soundscape of the birds captured in the vicinity. The server part manages the data regarding the audio files of the annotated birds with corresponding tags of bird classification, GPS coordinates and time. The sound scene data, are imported in the sound scenario constructor, which manages the processes of gestural interaction control, and the sound spatialization engine, which is responsible for placing the recordings in virtual space. 

\begin{figure}[!h]
    \centering
    \includegraphics[scale=0.35]{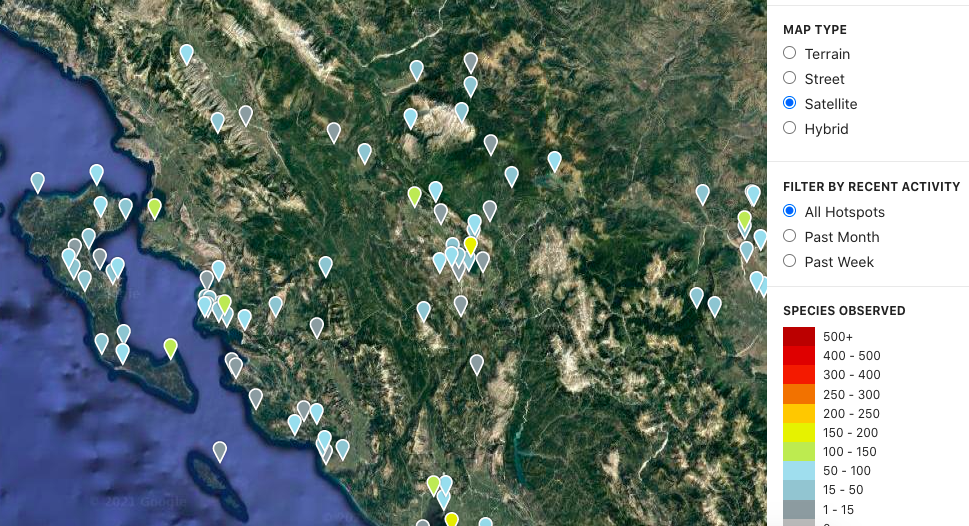}
    \caption{Indicative screen of the map taken from eBird project}
    \label{fig:my_label}
\end{figure}

\section{TECHNICAL IMPLEMENTATION}\label{sec:technical_implementation}
\label{sec:technical}

\subsection{Audio Capture}

The capture of the real acoustic environment is done by means of the sound recording features of the IoAuT device with. The user defines the recording option for the given scenario according to the available equipment. The available options include mono and binaural recording technology. The capture procedure also involves outputing the monitor audio to the default playback device. Mono recording uses modules that are typically available in almost all modern smartphones: the built-in microphone for detecting sounds of interest, and a set of headphones as playback acoustic equipment. The binaural recording option is performed using in-ear microphones embedded on a stereo headset like Sennheiser Ambeo Smart Headset\footnote{Sennheiser AMBEO Smart Headset-Mobile binaural recording headset, URL \url{https://en-de.sennheiser.com/in-ear-headphones-3d-audio-ambeo-smart-headset}, (Accessed on 03/19/2021)}. In both options, there is an optional gain control of the environmental monitoring system for personalized sound detection procedure. (fig.\ref{fig:TheConceptArchitecture}).

\subsection{Bird Call classification}
\begin{figure}[!t]
\centering
\includegraphics[trim={0cm 0.17cm 0.1cm 0cm},clip,width=\columnwidth]{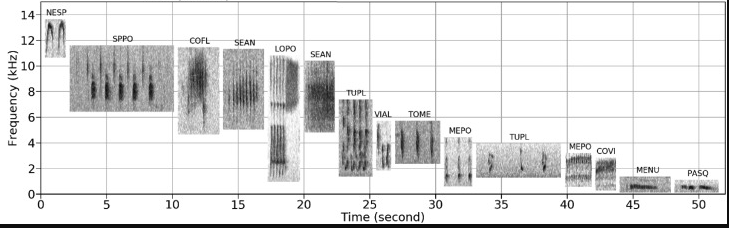}
\caption{Spectrograms of 14 distinct tropical birds from Puerto Rico. Taken from \cite{lebien2020pipeline}}
\label{fig:signature}
\end{figure}

Automatic bird sound classification plays an important role in monitoring and protecting biodiversity. Recent advances in machine listening and deep learning models for bird audio detection provide a novel way for improving bird call recognition to expert level. The fact that inter-species bird sounds exhibit such a distinctive spectral structure, motivated researches to employ typical hand-crafted time-frequency representations as an input to various deep learning models. The most prominent one is the usage of mel-band energies as the time-frequency representation, which is given as an input to 2D CNNs~\cite{lebien2020pipeline}. Figure~\ref{fig:signature} shows the distintive spectral structure of 14 tropical birds.

\begin{figure}[!t]
\centering
    \begin{subfigure}[b]{0.23\textwidth}
        \centering
        \includegraphics[width=\textwidth, height=3.5cm]{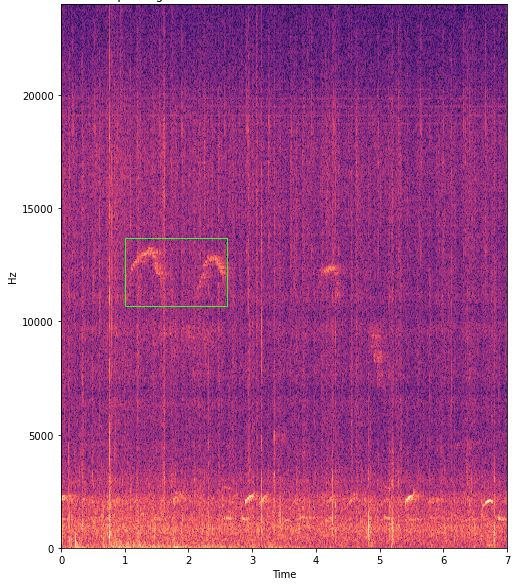}
       
    \end{subfigure}
    \begin{subfigure}[b]{0.23\textwidth}  
        \centering 
        \includegraphics[width=\textwidth, height=3.5cm]{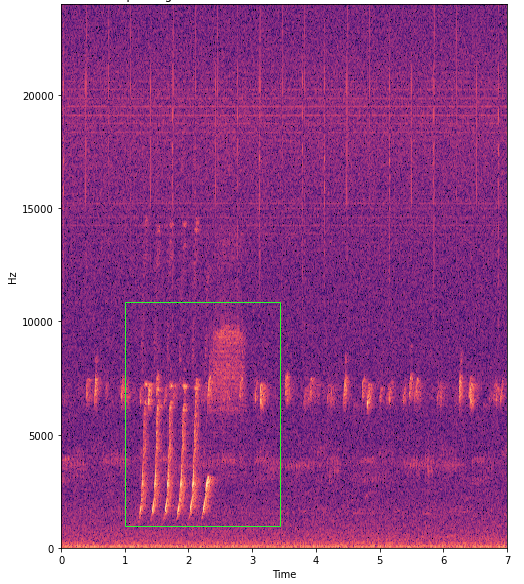}
  
    \end{subfigure}
    \vskip\baselineskip
    \begin{subfigure}[b]{0.23\textwidth}   
        \centering 
        \includegraphics[width=\textwidth, height=3.5cm]{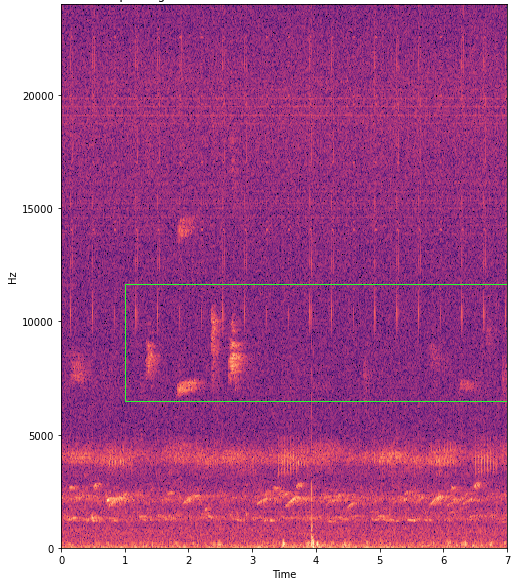}
   
    \end{subfigure}
    \begin{subfigure}[b]{0.23\textwidth}   
        \centering 
        \includegraphics[width=\textwidth, height=3.5cm]{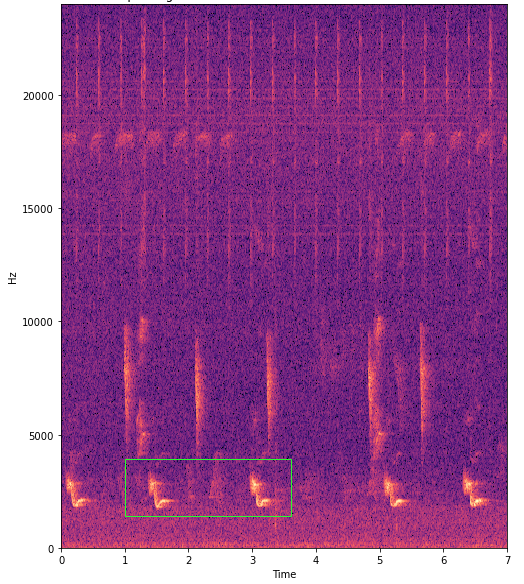}
    \end{subfigure}
    \caption{Time-frequency representation of 4 distinct bird sounds from ~\cite{lebien2020pipeline}, with an annotation bounding box created by an expert}\label{fig:birdsounds}
\end{figure}

The most recent example of a Bird audio classification model is BirdNet, introduced by Cornell University ~\cite{arif2020testing,kahl2021birdnet}. It involves a ResNet-like CNN model, containing 127 layers and 27 million parameters, and capable of identifying 984 North American and European bird species by sound. BirdNet achieves a Mean Average Precision (mAP) of $0.791$ for single-species recordings. However, typical CNNs cannot model long temporal dependencies that are usually needed in machine listening tasks, such as bird-audio detection~\cite{cakir2017convolutional,drossos:ijcnn:2020}. To address this issue, different published papers have adopted the Convolutional Recurrent Neural Network (CRNN) model~\cite{cakir2017convolutional,cakir:taslp:2017}. The CRNN model consists of a series of 2D CNNs, followed by Recurrent Neural Networks (RNNs) and a linear layer. A time-frequency representation of audio is given as an input to the a CRNN model, and the 2D CNNs learn time-frequency patterns according to the targeted task (e.g. bird audio detection). Then, the RNNs take as input the output of the CNNs, and focus on learning temporal patterns. Finally, the linear layer is fed the output of the RNNs and performs the classification. The CRNN model achieved high performance in DCASE Challenge tasks on bird-audio detection, ranking among the top 5 systems~\cite{cakir2017convolutional,adavanne2017stacked}.

\subsubsection{Workflow}
A typical workflow for training deep learning based models, with regard to bird audio/call recognition task, consists of the following steps:
\begin{enumerate}
    \item Data acquisition: The primary data sources are sounds recorded in the wild by experts who recognise the bird on the spot and keep extensive metadata about time and location and recording quality.
    \item Data pre-processing: Audio files are converted to time-frequency representations using short-time analysis and transform. Theses representations can be enhanced to expose the signature patterns of each bird. It is common practice to apply a digital band-pass filter to focus on a specific the frequency range. Mel-scale spectrograms are one obvious representation choice since they model the human hearing.
    \item Data labelling: For this task the expert examines raw audio files and locates the time frame where each bird can be detected. Furthermore,  time-frequency visualization can by also inspected manually to localize each bird visual pattern. The result can be like Figure~\ref{fig:birdsounds} where a green box is superimposed into the regions of each bird sound signature.
    \item Data augmentation: Domain-specific data jitter is used to account for unforeseen variations of real-world samples. This is imperative due to the huge acoustic domains shift  between studio and soundscape recordings. This data augmentation can be performed online during training both on the audio domain and on the time-frequency visualization.
    \item CNN architecture setting: The common procedure is to select a CNN architecture and train it, by minimizing an appropriate error function and using the labelled data. The input of the model depends on how many seconds of the input audio is needed to classify at once and the density of bird calls in our recordings. There are two major modeling options: i) clip-wise annotation/inference, where the model classifies a bunch of seconds at once, and ii) frame-wise annotation/inference where the model outputs prediction for each discrete time step (e.g. milliseconds). In practice, a validation set is used to control model over-fitting, following typical machine learning protocols and processes (e.g. early stopping). 
    \item Inference: Having trained a model on many annotated audio samples we are ready to deploy it to new sounds. During the inference phase, the input audio is subject to the same pre-processing steps as the audio signals used for training, ie. re-sampling, filtering, and converted into time frequency representation, before it is given as an input to the model.
\end{enumerate}

\subsubsection{Delivering the model} 
The trained model that results from the workflow of the previous section can be delivered to our application in two ways: i) recognition as a service where a back-end server (powered by GPU) is used to accept audio chunks, perform the pre-processing and the inference, and report the classification result and ii) recognition on the edge-device, where all process is performed on the user side. Each serving paradigm has pros and cons. Recognition as a service needs network access and a powerful back-end server, whereas recognition on the edge can accommodate relatively small sized models and has no network requirements.

\subsection{GIS based repository}
Since bird call detections will be accompanied by GPS based geographical coordinate and time stamp, all detection data can be presented in the form of cartographic GIS data model. This will allow users to easily locate their findings and the findings of others. GIS data can be aggregated by zoom levels and give finer grain detection the further the user zooms in. The basic user interface screen could be a map with overlaid information about the user's current geo-location and  already existing bird call detection. A side effect of this online collaborative GIS repository would be the extraction of migratory journeys of species of bird as well as the abundance of specific species.

\subsection{Augmented Reality Audio}

The ARA environment consists of a real and a virtual acoustic component with the real sound recording being mixed with the spatial reproduction of the classified captured birds' sounds into a pseudoacoustic environment, a mix that needs to take place as seemlessly as possible. The importance of a dynamic ARA mix of the gain difference between the real and the virtual acoustic environment compared to the static mix gain in legacy ARA mix models \cite{karjalainen2008augmented} has been pointed out in \cite{moustakas2019augmented}. The comparison of the legacy and the adaptive ARA mix model has shown that the latter demonstrated significantly better performance in terms of auditory perception \cite{moustakas2020adaptive}. Thus, in the proposed framework, we employ this dynamic and adaptive ARA mixing strategy that focuses on the impact of dynamic fluctuations of the real and the virtual environment to acoustic perception, taking in consideration acoustic phenomena, such as auditory masking.

Furthermore, the location awareness in ARA systems refers to the capability of a device to determine its location in terms of coordinates through active or passive human-computer interaction. Several ARA works have shown the necessity to utilize spatialisation techniques, in order to combine data extracted from location awareness systems with virtual sound sources \cite{mynatt1997audio,vazquez2012auditory}. Our proposed system includes a spatialisation module for positioning the 3d virtual sound sources, a set of sensors including gyroscope, accelerometer, and GPS, that facilitates gestural interaction, and a headset for reproducing the augmented acoustic environment. With this setup users are free to move their head in both horizontal and median plane, and listen to the entire 3d acoustic space, while transmitting their location, movement, and gestural activity to the system's engine.

\section{CONCLUSION}\label{sec:conclusions}

We have presented a framework for implementing a Citizen Science project in a Smart City environment, and enriching it with game elements using Internet of Sound technologies. The aim of our inter-disciplinary approach is to seek ways to enhance the public's motivation for participation, engagement in the experience, and deeper understanding of the subjects and processes at hand. We focused on Bird Call Recognition and Monitoring collaborative activities, in which participants can report and study the state and flux of the urban aerial ecosystem in a playful manner. We propose the use of game elements including quests, points and badges of progress, narratives, and structured levels, which agree with modern learning theories, such as Problem-based, Experiential, and Constructivist Learning. User interaction relies on IoAuT mechanisms including recording and submitting audio data, and exploring the augmented environment through GIS-related navigation and gestural performance with the mobile device. Sound classification takes place through a CNN network, and the augmented soundscape is constructed by an adaptive ARA mixing system.

A common core aspect of Citizen Science and Smart Cities is to focus on citizens’ problems and needs. The users of our proposed system are seen, on the one hand, as active units that exploit the benefits of the enhanced world around them, and, on the other, as interconnected members of the community that collaborate with each other to improve that world. Smart Cities facilitate a safe environment to observe, reflect, and experiment, whereas Citizen Science provides with specific problems to solve and thus contribute to the scientific community. Our proposed framework is currently in a pre-test stage; it needs to be tested in terms of its efficiency to address the intended objectives. We hope that it will serve as future reference towards enhancing the appeal of Citizen Science to the involved stakeholders, and providing novel ways to realize personal and collective interactive experiences based on a network of audio devices able to collect, evaluate, process and distribute acoustic data in urban environments.    

\section*{ACKNOWLEDGEMENT}
Correspondence should be addressed to Emmanuel Rovithis. K. Drossos is supported by the European Union’s Horizon 2020 research and innovation programme under grant agreement No 957337, project MARVEL.

\bibliography{aes2e.bib}
\bibliographystyle{aes2e.bst}

\end{document}